\documentclass[aps,prb,twocolumn,showpacs,floatfix,amsmath]{revtex4}

\usepackage{graphicx}
\begin{document}


\title{Quantum and classical resonant escapes of a strongly-driven Josephson junction}

\author{H. F. Yu, X. B. Zhu,  Z. H. Peng, W. H. Cao, D. J. Cui, Ye
Tian,\\ G. H. Chen, D. N. Zheng, X. N. Jing, Li Lu, and S. P. Zhao}

\affiliation{Beijing National Laboratory for Condensed Matter
Physics, Institute of Physics, Chinese Academy of Sciences, Beijing
100190, China}

\author{Siyuan Han}

\affiliation{Department of Physics and Astronomy, University of
Kansas, Lawrence, Kansas 66045, USA}


\begin{abstract}

The properties of phase escape in a dc SQUID at 25 mK, which is well
below quantum-to-classical crossover temperature $T_{cr}$, in the
presence of strong resonant ac driving have been investigated. The
SQUID contains two Nb/Al-AlO$_{x} $/Nb tunnel junctions with
Josephson inductance much larger than the loop inductance so it can
be viewed as a single junction having adjustable critical current.
We find that with increasing microwave power $W$ and at certain
frequencies $\nu $ and $\nu $/2, the single primary peak in the
switching current distribution, \textrm{which is the result of
macroscopic quantum tunneling of the phase across the junction},
first shifts toward lower bias current $I$ and then a resonant peak
develops. These results are explained by quantum resonant phase
escape involving single and two photons with microwave-suppressed
potential barrier. As $W$ further increases, the primary peak
gradually disappears and the resonant peak grows into a single one
while shifting further to lower $I$. At certain $W$, a second
resonant peak appears, which can locate at very low $I$ depending on
the value of $\nu $. Analysis based on the classical equation of
motion shows that such resonant peak can arise from the resonant
escape of the phase particle with extremely large oscillation
amplitude resulting from bifurcation of the nonlinear system. Our
experimental result and theoretical analysis demonstrate that at
$T\ll T_{cr}$, escape of the phase particle could be dominated by
classical process, such as dynamical bifurcation of nonlinear
systems under strong ac driving.

\end{abstract}


\maketitle

\section{Introduction}

Devices based on Josephson junctions are not only the key elements
for the realization of superconducting qubits \cite{makh01} but also
testbeds for the study of macroscopic quantum phenomena and
nonlinear dynamics. In these studies, response of the device to
resonant or near-resonant microwave fields provides insights to
understand their behavior. Numerous works have been reported in the
literature, such as the demonstration of energy level quantization
in current biased junctions
\cite{martinis85,clarke88,wallraff03,bauch06,xu03} and Rabi
oscillations in phase qubits. \cite{xu03,yu02,martinis02} These
experiments are usually performed at low temperatures ($\sim $ 20
mK) in relatively weak microwave field. Under such conditions, the
junctions behave quantum-mechanically and the phase particle may
escape from the potential well by macroscopic quantum tunneling
(MQT) from ground state and/or via photon-assisted tunneling (PAT)
from excited states, depending on the bias current $I$ and the
microwave frequency $\nu$. When the switching current distribution
$P(I)$ of a current biased junction is measured, MQT from ground
state manifests as a peak, which we call the primary peak, centered
slightly below the critical current $I_c$ of the junction. In the
case of the PAT process, a resonant peak at a slightly lower current
$I_{res}$ could be observed in addition to the primary peak in
$P(I)$.

The phase escape of a Josephson junction at low temperatures
exhibits a number of interesting phenomena as the microwave power
$W$ increases, which can be described as follows. (1) As mentioned
above, when the frequency $\nu$ matches the bias current dependent
energy level spacing between the ground and the first excited states
at $I_{res}$ that is close to $I_{c}$ and $W$ is low, the excited
state population $\rho_{1}$ will be enhanced but remains $\ll$ 1.
This results in a higher escape rate $\Gamma \simeq (1-\rho
_{1})\Gamma _{0}+\rho _{1}\Gamma _{1}>\Gamma _{0}$ at $I_{res}$
since $\Gamma _{1}\sim 10^{3}\Gamma _{0}$. Consequently, a resonant
peak at $I_{res}$ appears in $P(I)$. Because $W$ is low, the
system's potential and level spacing are essentially unperturbed by
microwave field, as reflected in the unchanged position of the
primary peak. (2) When $\nu $ corresponds to the level spacing at a
resonance current $I_{res}$ that is well below $I_{c}$ and the
microwave field is weak, there will be no resonant peak since
$\Gamma _{1}$ is still too low to observe at $I_{res}$. In this
case, as $W$ is increased, the position of the primary peak will
move continuously to lower bias currents due to the suppression of
the effective potential barrier. \cite{fistul03,fistul00,sun08} This
suppression is particularly strong at resonance, leading to a
significant increase of $\Gamma _{1}$ and thus a visible resonant
peak. (3) If $\nu $ matches a fraction of the level spacing and
satisfies $2\pi \nu m=\omega _{0n}$, where $\hbar \omega _{0n}$ is
the level spacing between the ground state and the $n$th excited
state and $m$ is a small positive integer, the multiphoton PAT
process occurs and the corresponding resonant peak appears. (4) When
$W$ further increases, the system's nonlinear characteristics such
as nonlinear resonance, bifurcation, and chaotic dynamics may
emerge.

The resonant peak has also been observed at higher temperatures
({\it e.g.} 4.2 K) in the classical regime where thermal activation
over the top of the barrier dominates.
\cite{devorte84,mao07,tonseca86,rotoli07} In this regime, when the
driving frequency $\nu$ matches the junction's bias current
dependent plasma frequency $\omega _{p}(I)$, the classical resonance
occurs. The oscillation amplitude of the phase particle in the
potential well and thus the escape rate can be greatly enhanced.
Recently, such process and their classical interpretation have been
studied in extended temperature range and have received much
attention due to a series of works by Gronbech-Jensen and coworkers.
\cite{jensen04a,jensen04,jensen06,jensen05,marchese06} These authors
demonstrated that previously quantum-mechanically explained single-
or multi-photon resonance, the effective barrier suppression, and
the Rabi oscillation can also be understood from the classical point
of view. Interpretations on both the classical and quantum bases are
also reported \cite{jensen05,marchese06,claudon07,strauch07} for the
results involving multiphoton, multilevel Rabi oscillation, and ac
Stark shift at high $W$.
\cite{claudon04,claudon07,strauch07,schuster05} It is found that at
high $W$, the predictions of the classical and quantum pictures may
converge. \cite{claudon07} These results indicate that the Josephson
junction system can have a {\it dual} character, classical and
quantum-mechanical, when a microwave field is applied.

In this paper, we investigate the switching current distribution and
the escape rate of a dc SQUID as a function of microwave field
strength at low temperatures. The SQUID contains two
Nb/Al-AlO$_{x}$/Nb junctions with Josephson inductance much greater
than the loop inductance so it behaves as a single junction with a
tunable critical current. We found that as microwave power $W$ is
increased while frequency $\nu $ is kept constant, the primary peak
of $P(I)$ first shifts toward lower bias current and then a resonant
peak develops. This result may find a classical explanation,
\cite{jensen04,jensen06,sun08} but here we will show instead that it
is also in accordance with the situation (2) discussed above. We
further found that the effect of two-photon process on escape rate
can be observed in this situation. Moreover, as $W$ further
increases, the primary peak will gradually disappear and the
resonant peak grows into a single one while shifting continuously to
the lower bias current. At even higher $W$, a second resonant peak
appears, which could locate at a very low bias current depending on
the value of $\nu $. Analysis based on the classical equation of
motion indicates that such double-resonant-peak structure with a
large separation between the two peaks at high $W$ can originate
from the resonant escape of the phase particle having a very large
oscillation amplitude resulting from bifurcation of the nonlinear
system driven by a strong microwave field.
\cite{landau,sidd05,manu07}

\section{Experimental techniques and results without microwave radiation}

\begin{figure}[t]
\includegraphics[width=0.48\textwidth]{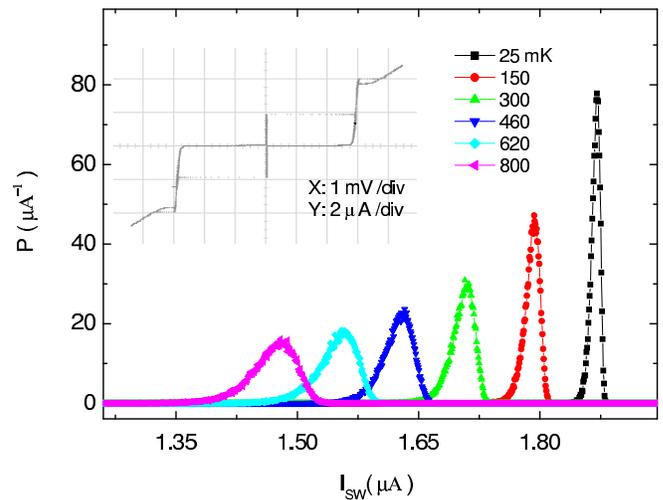}
\caption{(Color online) Switching current distribution $P(I)$ at
several temperatures between 25 and 800 mK for the sample used in
this work. The inset shows the oscilloscope $I$-$V$ image at 25 mK.}
\end{figure}

The sample used in this work was a dc SQUID containing two
Nb/Al-AlO$_{x}$/Nb junctions, which was prepared at Stony Brook
using a self-aligned process. \cite{chen04} Each junction had a
diameter of $1.27$ $\mu $m (corresponding to an area of $1.27$ $\mu
$m$^{2}$) and the SQUID loop had an area of $220$ $\mu $m$^{2}$. The
device's oscilloscope $I$-$V$ trace at $25$ mK is displayed in the
inset of Fig.~1. Taking $I_{c}/2$ $\sim $ $1$ $\mu $A, we find the
Josephson inductance of $330$ pH for each junction, which is much
larger than the SQUID loop inductance of $\sim$ 30 pH. Hence, the
SQUID can be viewed as a single junction having a maximum critical
current $I_{c}$.

According to the RCSJ model, \cite{rcsj} the dynamics of a single
junction biased at $I$ can be described by a fictitious phase
particle with position $\varphi $, mass $M=C(\Phi _{0}/2\pi )^{2}$,
and friction coefficient $1/RC$ moving in a washboard potential
\begin{equation}
U(i,\varphi )=-E_{J}(i\varphi +\cos \varphi ),  \label{U_JJ}
\end{equation}
\noindent in which $\Phi _{0}=h/2e$ is the flux quantum,
$E_{J}=I_{c}\Phi _{0}/2\pi $, and $i\equiv I/I_{c}$. $C$ and $R$ are
the capacitance and shunt resistance of the junction, respectively.
The plasma frequency of the junction is given by $\omega _{p}=\omega
_{0}(1-i^{2})^{1/4}$, where $\omega _{0}=\sqrt{2\pi I_{c}/C\Phi
_{0}}$. For $i<1$, the phase particle may escape from the potential
well either by thermal activation (TA) or by macroscopic quantum
tunneling, resulting in the junction's switching from the zero to
the finite voltage state. Crossover from MQT to TA is known to occur
at $T_{cr}=\hbar \omega _{p}[(1+\alpha ^{2})^{1/2}-\alpha ]/2\pi
k_{B}$, where $\alpha =1/2Q$, and $Q=\omega _{p}RC$ is the quality
factor. The escape rate in the TA regime can be found from Kramers'
formula: \cite{kram40}
\begin{equation}
\vspace{1pt}\Gamma _{t}=\frac{\omega _{p}}{2\pi }a_{t}\exp \left(
-\frac{\Delta U}{k_{B}T}\right)  \label{TA}
\end{equation}
\noindent where $a_{t}=4/(\sqrt{1+Qk_{B}T/1.8\Delta U}+1)^{2}$ is a
damping dependent factor, and in the MQT regime from: \cite{cald81}
\begin{equation}
\Gamma _{q}=\frac{\omega _{p}}{2\pi }a_{q}\exp \left[
-\frac{7.2\Delta U}{\hbar \omega _{p}}\left( 1+\frac{0.87}{Q}\right)
\right] \label{MQT}
\end{equation}
\noindent where $a_{q}\simeq \lbrack 120\pi (7.2\Delta U/\hbar
\omega _{p})]^{1/2}$. In the above expressions, the barrier height
is given by
\begin{equation}
\Delta U(i)=2E_{J}[\sqrt{1-i^{2}}-i\arccos (i)]~.
\end{equation}

\begin{figure}[t]
\includegraphics[width=0.45\textwidth]{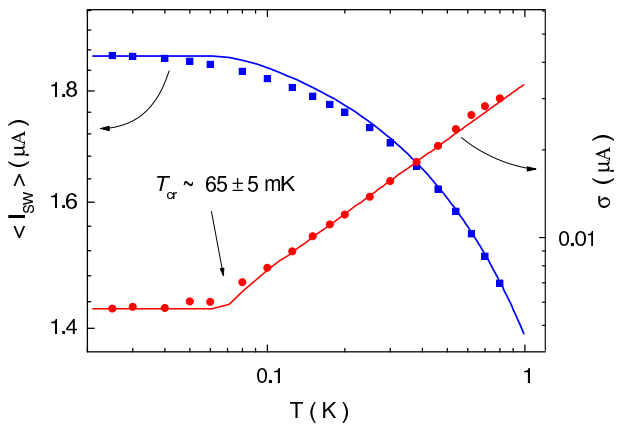}
\caption{(Color online) Temperature dependence of the mean
$<I_{sw}>$ and standard deviation $\protect\sigma$ of the measured
$P(I)$ (symbols). Solid lines are the predictions from TA and MQT
theories using the parameters listed in Table I. The experimental
MQT-to-TA crossover temperature $T_{cr}=65\pm 5$ mK is indicated.}
\end{figure}

\begin{figure*}[t]
\centering
\begin{minipage}[c]{0.7\textwidth}
\scalebox{0.6}[0.6]{\includegraphics{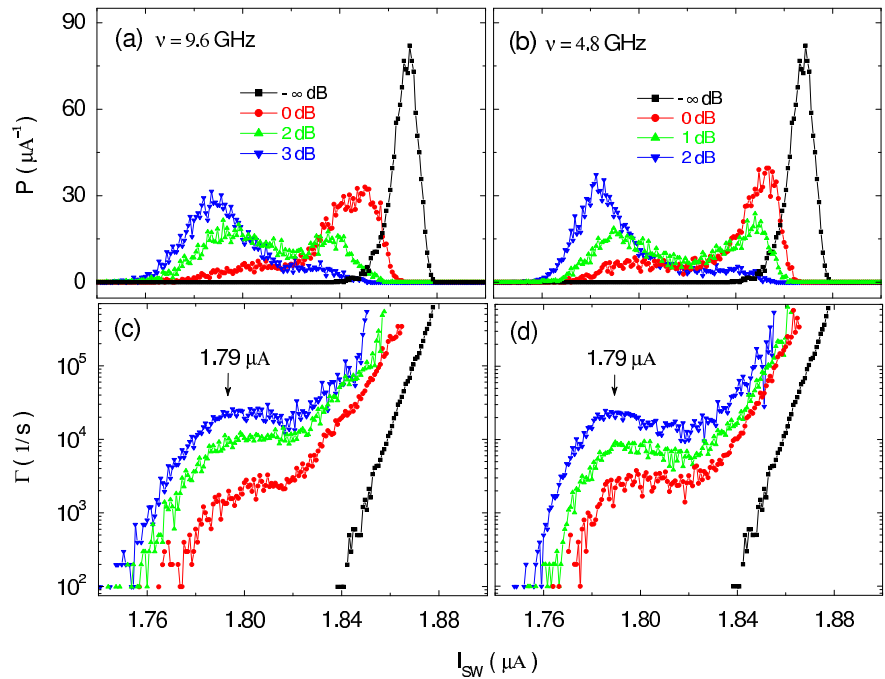}}
\end{minipage}
\centering \caption{(Color online) Switching current distribution
$P(I)$ and escape rate $\Gamma(I)$ at 25 mK for several microwave
powers as indicated. Power levels are relative with the curves where
resonant peaks start to appear set to 0 dB, and the two 0 dB curves
in (a) and (b) correspond to -10 dBm and -37 dBm, respectively.
Resonant peaks demonstrating resonant quantum phase escape can be
clearly seen. (a) and (c): $\nu$ = 9.6 GHz, single-photon process;
(b) and (d): $\nu$ = 4.8 GHz, two-photon process. The resonance
current $I_r$ is indicated by the arrows in (c) and (d).}
\end{figure*}

We measured switching current distribution $P(I)$ using the
time-of-flight technique, as described in our previous works.
\cite{li07,cui08} The results at some typical temperatures are shown
in Fig.~1. In our experiment, the bias current was ramped up at a
constant rate of $dI/dt$=0.4 mA/s with a repetition rate of 97 Hz.
To reduce statistical uncertainty each measured $P(I)$ contained
$5\times 10^{4}$ escape events. The sample was placed in a copper
box, which was anchored to the mixing chamber (MXC) of an Oxford
Kelvinox MX400 dilution refrigerator. To reduce the effects of
external noise on switching dynamics to a negligible level, a
trilayer $\mu $-metal shield was used and the circuit was filtered
by EMI filters at room temperature, cryogenic low-pass filters at 1
K pot, and copper powder microwave filters at MXC temperature. Since
noise could cause temperature dependence of the width of $P(I)$ to
flatten out at $T>T_{cr}$, which could be mistaken as the evidence
for MQT, we tested the noise level by applying a flux to the SQUID
loop to reduce its critical current to $I_c(\Phi)$ $\ll I_{c}(0).$
The result confirmed that noise from the environment and measurement
circuitry was negligible down to 25 mK. (The data presented in this
work were however obtained with maximum critical current $I_c$ =
$I_{c}(0)$ in the absence of applied magnetic field. The plasma
frequency at bias current $\sim$ 1.865 $\mu$A, which is the position
of maximum $P(I)$ at 25 mK in Fig.~1, is $\omega _{p}$ $\sim$ 10
GHz.)

\begin{table}[b]
\caption{Sample parameters.} \label{tab:table}
\begin{ruledtabular}
\begin{tabular}{cccc}
Junction area&$I_c$ &C  &R \\
\hline
2$\times$1.27 $\mu$m$^2$&1.957 $\mu$A &620 fF &300 $\Omega$ \\
\end{tabular}
\end{ruledtabular}
\end{table}

Fig.~1 shows that as temperature decreases, $P(I)$ narrows and moves
to higher bias current. In Fig.~2, we show the temperature
dependence of the mean ($\langle I_{sw}\rangle $) and width ($\sigma
$) of $P(I)$ (symbols), together with the results calculated from
the TA and MQT theories (lines) using the standard conversion
between $\Gamma _{t,q}$ and $P(I)$. \cite{fult74} From the fit, the
sample's parameters $I_{c}=1.957$ $\mu$A, $C=620$ fF, and $R=300$
$\Omega $ were obtained. \cite{rem} In Fig.~2, we can see that in
the temperature range between 1 K and $T_{cr}$ $\simeq $ $65$ mK,
$\sigma $ decreases with decreasing temperature indicating that TA
is the dominant escape mechanism. In this temperature range, $P(I)$
depends weakly on $C$ and $R$, so $I_{c}$ can be determined by
fitting $P(I)$ using the TA theory starting with a reasonable
estimate of $C$ and $R$ values. The values of $C$ and $R$ were then
determined using $P(I)$ obtained at $T\ll T_{cr}$ because the
exponent of MQT rate is dependent on $C$ and $R$. In Table I, we
summarize the sample's key parameters. Taking the $P(I)$ peak
position at 25 mK in Fig.~1 again, we find that the crossover
temperature calculated from these parameters is $T_{cr}$ = 62.4 mK,
which is in good agreement with the experimental value indicated in
Fig.~2.

\section{Results with microwave radiation}

Fig.~3(a) shows the switching current distribution $P(I)$ at $25$ mK
in the microwave field with fixed frequency $\nu $ =9.6 GHz and four
different power levels. (In this work, we always use relative power
levels by normalizing to the power level at which the resonant peak
first appears.) Fig.~3(c) shows the corresponding escape rate
$\Gamma $. It is seen that as $W$ increases, the primary peak of
$P(I)$ shifts to lower bias current and then a resonant peak
develops. Further increasing $W$ causes the primary peak to
disappear and resonant peak to grow into a single one (Results at
higher $W$ are presented in Fig.~4 below). This result, though
similar to situation (1) discussed in Section I, can be interpreted
along a quantum-mechanical analysis by Fistul {\it et al.}
\cite{fistul03} since the primary peak demonstrates a clear leftward
shift resulting from effective barrier suppression. \cite{rem1}

According to the discussion in Ref.~\onlinecite{fistul03}, the
condition $(2\pi \nu /\omega_0)^{5}$ $\leq $ $\hbar \omega_0/E_{J}$
should be satisfied in order for the process described in situation
(1) to occur. The physics behind this is that the microwave-induced
excited-level population times the escape rate of the level should
be large. Taking the junction's parameters in Table I, we find
$(2\pi \nu /\omega_0)^{5}$ $\sim $ $0.088$ and $\hbar
\omega_0/E_{J}$ $\sim $ $0.016$ so the condition is clearly not
satisfied. On the other hand, we calculated the energy levels, and
the level escape and transition rates of the system using the
approach described in Ref.~\onlinecite{kopi88}. The total escape
rate $\Gamma _{LO}$ was then obtained from the stationary solution
of the Master equation as first considered by Larkin and
Ovchinnikov. \cite{xu03,kopi88,lo86} We found that for the data
without microwave radiation (-$\infty$ dB) in Figs.~3(a) and (c),
there were three levels in the potential well and the calculated
$\Gamma _{LO}$ agreed well with the experimental result. If we
assume an unsuppressed potential barrier in the microwave field and
use $I_{res}$ =1.79 $\mu $A at the resonant peak, we would have 6
energy levels in the well. In this case, even assuming all
population at the first excited level would still lead to $\Gamma
_{LO}$ about 3 to 4 orders smaller than the measured escape rate.
From these results, it is clear that a barrier suppression resulting
in a reduction of one or two energy levels in the well is necessary
to account for the experimental data in Fig.~3.

The calculated level spacing $\omega _{01}/2\pi $ between the ground
and first excited states at $I_{res}$ =1.79 $\mu $A, which can also
be estimated from $\omega _{01}=\omega _{p}(1-5\hbar \omega
_{p}/36\Delta U)$, is $\sim $ 9.6 GHz, which exactly matches the
frequency of the applied microwave. This result demonstrates that
the quantum explanation of the resonant phase escape process is
applicable. In Figs.~3(b) and (d), we show the corresponding results
when the microwave frequency is reduced to $\nu $ $=(\omega
_{01}/2\pi )/2=4.8$ GHz. Similar behavior can be seen, which is
consistent with the quantum picture of the resonant phase escape as
a result of the two-photon process. \cite{additional}

As $W$ increases, the primary peak in Fig.~3 decreases and
eventually disappears while the resonant peak grows into a single
one. The data taken at $\nu =9.6$ GHz are plotted in Fig.~4. The
most interesting feature of the data in this figure is that a second
resonant peak develops when $W$ is further increased. \cite{rem2}
Compared with the previous case of weak microwave field, the
separation between the second resonant peak and the first resonant
peak is much larger than that between the first resonant peak and
the primary peak. Furthermore, the separation is found to depend
strongly on the microwave frequency $\nu $. In Fig.~5, we show the
$W$ dependence of $P(I)$ at $25$ mK when $\nu $ is changed to $7$
GHz which is largely detuned from the level spacing $\omega
_{01}/2\pi \simeq 9.6$ GHz in the vicinity of the bias current where
the primary peak is located. Therefore, resonant peak originated
from PAT ({\it e.g.} the peaks just below the primary peak in
Figs.~3 and 4) would not occur at low microwave power. As $W$
increases, the system gradually enters the strong driving regime.
The primary peak is observed to shift leftwards continuously until a
resonant peak at a much lower bias current $\sim $ $0.85$ $\mu $A
develops (note the different horizontal scales of Figs.~4 and 5).
Such distinct results could originate from Josephson bifurcation,
\cite{sidd05,manu07} a phenomenon common to strongly-driven
nonlinear dynamic systems, as discussed below.

\section{Discussion based on the classical nonlinear dynamics}

It should be pointed out that the number of levels in the potential
well at the second resonant peak (at $\sim $ $1.5$ $\mu $A) in
Fig.~4 and the corresponding peak at $\sim $ $0.85$ $\mu $A in
Fig.~5 would be about $20$ and over $50$, respectively. The presence
of so many energy levels during resonant escape, provided a dramatic
barrier suppression did not occur, necessitates a many-photon
process from the quantum mechanical point of view or motion with
extremely large amplitude oscillations from the perspectives of
classical dynamics. Since it is not clear whether the treatment of
Fistul {\it et al.} \cite{fistul03} is applicable to such case we
provide an explanation based on the classical description of dynamic
bifurcation in nonlinear systems for the experimental results
presented in Figs.~4 and 5. Our analysis shows that the experimental
results obtained at high $W$, namely the double-peak structure with
large peak separation, can be caused by escape of the phase particle
having very large oscillation amplitude resulting from bifurcation
of the nonlinear system driven by strong microwave field.
\cite{landau,sidd05,manu07}

\subsection{Nonlinear bifurcation phenomenon: Bistable oscillations with
different amplitudes}

\begin{figure}[t]
\includegraphics[width=0.48\textwidth]{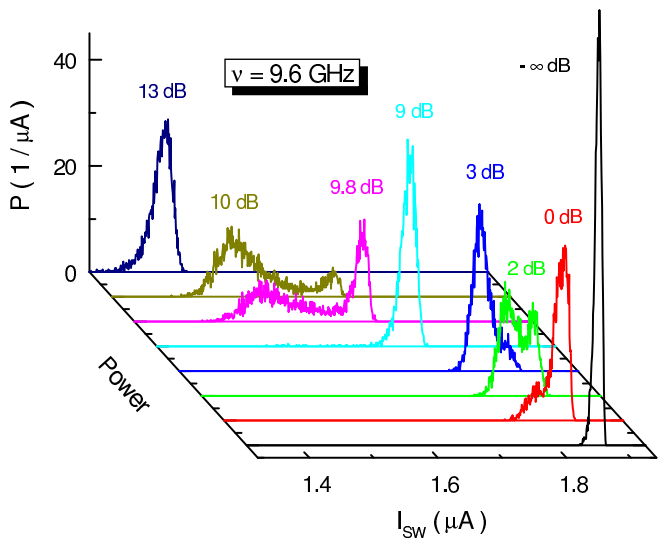}
\caption{(Color online) Switching current distribution $P(I)$ at 25
mK with different relative microwave power $W$. The 0 dB curve
corresponds to a power level of -10 dBm. The microwave frequency is
$\protect\nu$ = 9.6 GHz. It is seen that as $W$ increases, the
double-peak structure appears and disappears twice. The appearance
of the two resonant peaks can be explained by quantum resonant
escape and nonlinear classical resonant escape, respectively.}
\end{figure}

We start with the normalized equation of motion for the phase
particle:
\begin{equation}
\ddot{\varphi}+\frac{1}{Q}_{0}\dot{\varphi}+\sin \varphi
=i+i_{mw}\cos (\Omega \tau )~,
\end{equation}
\noindent where $\tau =\omega _{0}t$, $Q_{0}=\omega _{0}RC$,
$i_{mw}=I_{mw}/I_{c}$ is the normalized ac current induced by
microwave field, and $\Omega $ is the microwave frequency $\nu $
normalized to $\omega _{0}/2\pi $. At low temperatures ($T=25$ mK)
and strong microwave field, the effect of noise current $i_{n}$ due
to thermal fluctuations on escape rate is negligible and thus
omitting $i_{n}$ in Eq.~(5) is justified.

\begin{figure}[t]
\includegraphics[width=0.48\textwidth]{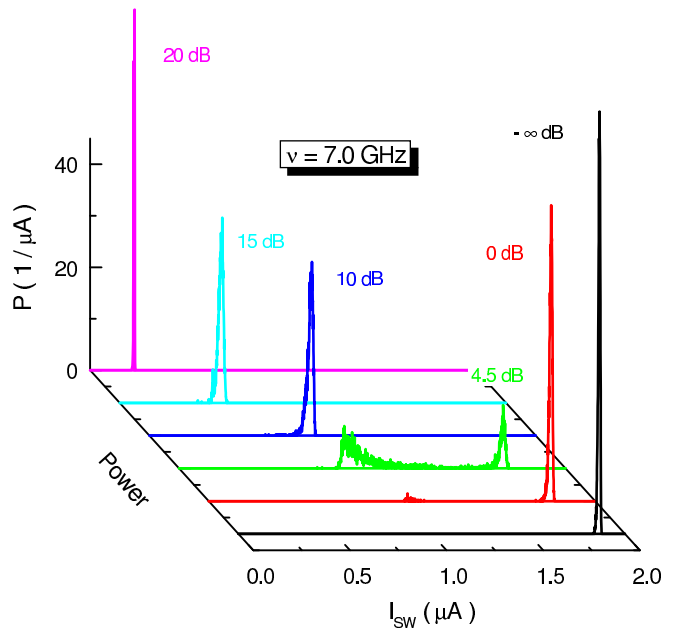}
\caption{(Color online) Switching current distribution $P(I)$ at 25
mK with different relative microwave power $W$. The 0 dB curve
corresponds to a power level of -15 dBm. The microwave frequency is
$\nu$ = 7 GHz. In contrast to the results in Fig.~4, the double-peak
structure appears and disappears only once. Here the resonant peak
results from the nonlinear classical resonant escape of the phase
particle.}
\end{figure}

Following the analysis of Refs.~\onlinecite{jensen05} and
\onlinecite{fistul03}, we use a monochromatic ansatz for the
solution of the unperturbed Eq.~(5):
\begin{equation}
\varphi =\varphi _{0}+\phi (\tau )
\end{equation}
\noindent with
\begin{equation}
\varphi _{0}=\arcsin (i)~.
\end{equation}
\noindent Inserting this ansatz into Eq.~(5) and expanding $\sin \phi $ and $%
\cos \phi $ up to the third order yield: \cite{rem3}
\begin{equation}
\ddot{\phi}+\frac{1}{Q}_{0}\dot{\phi}+\Omega _{1}^{2}\phi +a\phi
^{2}+b\phi ^{3}=i_{mw}\cos (\Omega \tau )~,
\end{equation}
\noindent in which $\Omega _{1}=(1-i^{2})^{1/4}$, $a=-i/2$, and $b=-\sqrt{%
1-i^{2}}/6$. When the applied microwave frequency $\Omega $ is close
to the resonance, we can write $\Omega =\Omega _{1}+\delta \Omega $,
where $\delta \Omega $/$\Omega$ $\ll$ 1, and make the single-mode
approximation:
\begin{equation}
\phi =A\sin (\Omega \tau +\eta )~,
\end{equation}
\noindent where $\eta $ is a phase constant and $A$ is amplitude of
the driven oscillation. Inserting this into Eq.~(8), one obtains the
following relation: \cite{landau}
\begin{equation}
A^{2}(\delta \Omega -\chi A^{2})^{2}+\frac{A^{2}}{4Q_{0}^{2}}=\frac{%
i_{mw}^{2}}{4\Omega _{1}^{2}}~,
\end{equation}
\noindent where $\chi =(3b/8\Omega _{1}-5a^{2}/12\Omega _{1}^{3})$.

\begin{figure}[t]
\includegraphics[width=0.45\textwidth]{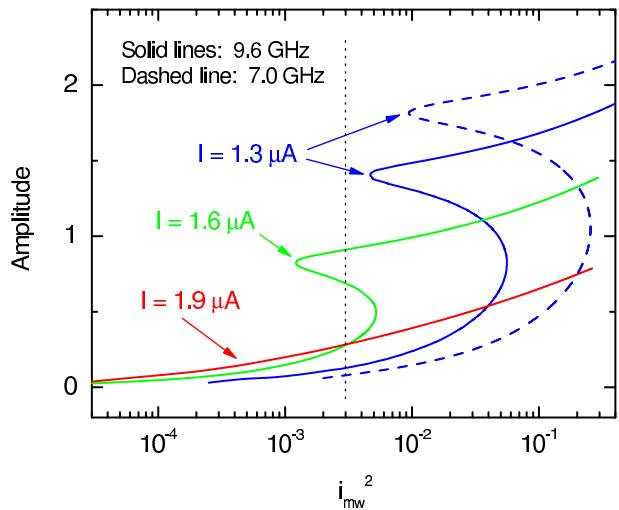}
\caption{(Color online) Calculated oscillation amplitude versus
$i_{mw}^2$ ($\propto$ microwave power $W$) of the phase particle for
different bias current $I$ and frequency $\nu$. The parameters in
Table I are used. The results show the familiar bifurcation at low
$I$ and high $W$, with two oscillation amplitudes $A_2$ $>$ $A_1$
(the middle one being unstable for a given curve).}
\end{figure}

From Eq.~(10), the amplitude $A$ as a function of microwave power
$W$ $\propto $ $i_{mw}^{2}$ and \textrm{bias current} $i$ can be
obtained since $\delta \Omega ,$ $\chi ,$ $Q_{0}$, and $\Omega _{1}$
only depend on $i$ and/or $W$ when microwave frequency $\Omega $ is
constant. Thus, we have
\begin{equation}
A=A(i,W)~.
\end{equation}
In Fig.~6, we present the calculated $A(i,W)$ for three indicated
bias current $i$ at $\nu =9.6$ GHz (solid lines) using the
parameters listed in Table I. The results show clearly the familiar
Josephson bifurcation phenomenon. \cite{sidd05,manu07} Namely, as
$W$ $\propto $ $i_{mw}^{2}$ increases, the system can be driven into
a regime where two dynamic states with different amplitudes of
oscillation $A_{2}$ $>$ $A_{1}$ exist, as shown schematically in
Fig.~7. In Fig.~6, one can see that in the range of $i_{mw}^{2}$
considered bifurcation does not occur at bias current $I=1.9$ $\mu
$A which is slightly below $I_{c}$ ($i\approx 0.971)$. As $I$
decreases, it starts to develop. For lower bias current, threshold
for bifurcation shifts to higher $W$ which is accompanied by a
greater amplitude $A_{2}$ at the threshold while $A_{1}$ has a slow
variation (the solid lines in Fig. 6). The dashed line in Fig.~6 is
the result obtained using $I=1.3$ $\mu $A and $\nu =7.0$ GHz. In
this case, $A_{2}$ is significantly larger when compared to the case
of same $I$ but higher driving frequency $\nu =9.6$ GHz. This result
will be shown to be consistent with, and can be used to explain the
changes from Fig.~4 to Fig.~5 when $\nu$ decreases.

\begin{figure}[t]
\includegraphics[width=0.43\textwidth]{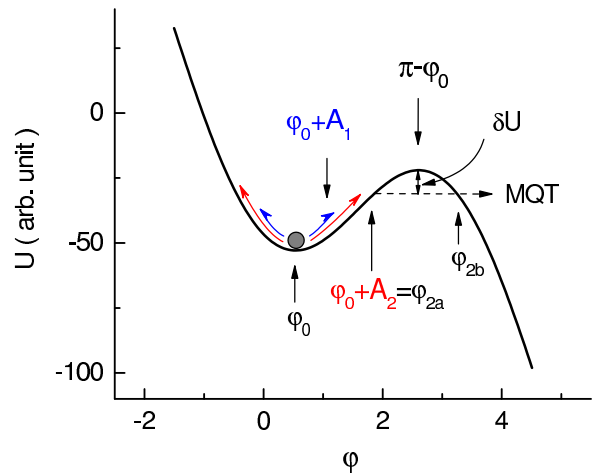}
\caption{(Color online) Washboard potential $U(\protect\varphi)$ of
Eq.~(1) with various phase values, remaining barrier height
$\protect\delta U$, and MQT process indicated.}
\end{figure}

\subsection{Resonant phase escape from bifurcated oscillation with the larger amplitude:
Explanation for the results at $\nu$ = 9.6 GHz}

We now show that the presence of $A_{2}$ resulting from bifurcation
at high microwave power can qualitatively explain the double-peak
structure with large peak separation as presented in Figs.~4 and 5.
We first consider the four curves with relative power levels of 9,
9.8, 10, and 13 dB in Fig.~4 with applied microwave frequency $\nu
=9.6$ GHz. In this case, all peaks are located at relatively high
bias current, so from the results presented in Fig.~6, the maximum
$A_{2}$ is seen to be around 1.5, which makes the approximation used
in deriving Eq.~(8) appropriate. \cite{rem3}

Since our measurement was performed at $25$ mK, escape of the phase
particle from the potential via thermal activation is negligible.
Hence we consider a classical resonantly-oscillating particle and
its MQT escape from the potential well. In general, the particle can
escape from both $A_{1}$ and $A_{2}$ states, thus the MQT rate, in
the absence of the barrier suppression, can be calculated using the
WKB method from:
\begin{eqnarray}
\Gamma _{WKB} &=&p_{1}\Gamma _{A_1}+p_{2}\Gamma _{A_2}~  \nonumber \\
&\simeq &p_{1}\nu ~\exp (-2S_{1}/\hbar )+~p_{2}\nu ~\exp
(-2S_{2}/\hbar )  \nonumber \\
&\simeq &p_{2}\nu ~\exp (-2S_{2}/\hbar ),~~~~~~~~~~~~~~~~~~
\label{WKB}
\end{eqnarray}
where $p_{1,2}$ are the probabilities of the phase particle in the
$A_{1}$ and $A_{2}$ states ($p_{1}+p_{2}=1$) and $S_{1,2}$ are the
corresponding path integral across the remaining tunnel barriers.
Because $S_{2}\ll S_{1}$ ($A_{2}\gg A_{1})$ tunneling rate from the
$A_{1}$ state is exponentially smaller and the escape is dominated
by the latter. In an explicit form, we can write down the action for
the phase particle in the $A_{2}$ state as:

\begin{equation}
S_{2}=\int\limits_{\varphi _{2a}}^{\varphi _{2b}}\sqrt{U(i,\varphi
)-U(i,\varphi _{2a})}d\varphi ~,
\end{equation}
in which $U$ is given by Eq.~(\ref{U_JJ}). The turning point at the
outer wall of the potential $\varphi _{2a}=\varphi _{0}+A_{2}$, the
point of escape $\varphi _{2b}$, and the corresponding path of MQT,
when the phase particle is in the $A_{2}$ state, are illustrated in
Fig.~7.

\begin{figure}[t]
\includegraphics[width=0.45\textwidth]{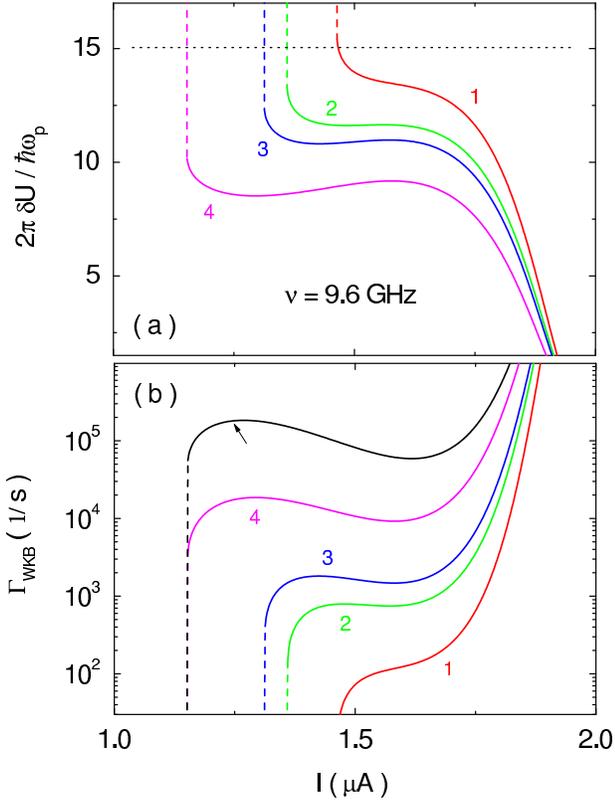}
\caption{(Color online) (a) Calculated $2\pi \delta U/\hbar
\omega_p$ in Eq.~(14) versus $I$ for the microwave frequency $\nu$ =
9.6 GHz. Curves labeled 1 to 4 have $i_{mw}^2$ = 2.5$\times$
10$^{-3}$, 3.7$\times$10$^{-3}$, 4.3$\times$10$^{-3}$, and
6.5$\times$10$ ^{-3}$. The parameters in Table I are used in the
calculations. The dotted line is the exponent $7.2\Delta
U(i_p)/\hbar \omega_p(i_p)$ in Eq.~(3) where $i_p$ is the value
corresponding to the primary peak without microwave radiation. (b)
MQT rate calculated from Eq.~(12) or (14) using $p_2$ = 10$^{-2}$
and the same $\nu$ and $i_{mw}^2$ parameters from curves 1 to 4 in
(a). The line pointed by an arrow has different $\nu$ = 9.4 GHz and
$i_{mw}^2$ = 6.8$\times$ 10$^{-3}$.}
\end{figure}

For the present experimental data, we usually have a deep potential
well $\Delta U$ [see Eq.~(4), for bias current well below $I_{c}$]
and the remaining barrier height $\delta U$ (see Fig.~7) for the
particle having $A_{2}$ oscillation amplitude is comparatively
small. In this case, it is straightforward to show that
Eq.~(\ref{WKB}) can be well approximated by
\begin{equation}
\Gamma _{WKB}\simeq p_{2}\nu \exp [-2\pi \delta U/\hbar \omega
_{p}]~, \label{WKB2}
\end{equation}
where $\delta U$ is given by
\begin{equation}
\delta U(i,W)=U(i,\pi -\varphi _{0})-U(i,\varphi _{0}+A_{2}(i,W))~.
\label{deltaU}
\end{equation}
Various phase locations in Eq.~(\ref{deltaU}) are indicated in
Fig.~7. In Fig.~8(a), we show the tunneling exponent $2\pi \delta
U/\hbar \omega _{p}$ of Eq.~(\ref{WKB2}) versus $I$ for $\nu =9.6$
GHz as $i_{mw}^{2}$ is increased from curve $1$ to curve $4$. Notice
that for each curve there exists a nearly vertical part, shown as
dashed lines, which corresponds to the sudden emergence of the
$A_{2}$ state. This can be understood by considering the case for
$i_{mw}^{2}=3\times 10^{-3}$ as indicated by the vertical dotted
line in Fig.~6. At $I=1.3$ $\mu $A, only the smaller amplitude
$A_{1}$ state exists. The amplitude $A_{1} $ would increase as $I$
increases. At a bias current $1.3$ $\mu $A $<I<1.6$ $\mu $A, the
much larger amplitude $A_{2}$ state emerges which results in the
sudden decrease of $\delta U(i)$ and thus a much larger tunneling
rate. In Fig.~8(a), it can be seen that the curve 1 shows a
monotonic dependence on $I$. Starting from the curve 2, however, a
local minimum develops. This behavior can in principle lead to the
double-peak structure in $P(I)$ as $i_{mw}^{2}$ is increased.

To compare with $\Gamma _{q},$ which is the MQT rate in the absence
of microwave radiation, we plot the exponent $7.2\Delta
U(i_{p})$/$\hbar \omega _{p}(i_{p})$ in $\Gamma _{q}$ [see
Eq.~(\ref{MQT}) considering $Q$ $\sim $ 10] as a dotted line in
Fig.~8(a), where $i_{p}$ corresponds to the position of the primary
peak. Notice that the line sits well above the wide shoulder part of
the solid curves resulting from the $A_{2}$ state of oscillation.
Since tunneling rate depends exponentially on $\delta U/\hbar \omega
_{p}$ our data imply $p_{2}\sim \exp [2\pi \delta U/\hbar \omega
_{p}-7.2\Delta U(i_{p})/\hbar \omega _{p}(i_{p})]\ll 1.$ Namely, the
particle spends more time in the $A_{1}$ state than in the $ A_{2}$
state. \cite{rem4}

In Fig.~8(b), we show $\Gamma _{WKB}$ calculated from
Eq.~(\ref{WKB}) or (\ref{WKB2}) using $p_{2}$ $=10^{-2}$ and the
same set of parameters as those in Fig.~8(a). The choice of $p_{2}$
around $10^{-2}$ brings the escape rate $\Gamma _{WKB}$ for the
system in the $A_{2}$ state of oscillation into the experimentally
observable range of $10^{2}$ to $10^{6}$ sec$^{-1}$ in Figs.~3(c)
and (d), which is determined by the bias current sweeping rate of
$dI/dt=0.4$ mA/s. It should be noticed that although the calculated
bias current dependence of $\Gamma _{WKB}$ in Fig.~8(b) looks
similar to the data presented in Figs.~3(c) and (d) as the microwave
power increases, the non-monotonic part of $\Gamma _{WKB}(I)$ covers
a much larger range of bias current. Furthermore, unlike the data in
Figs.~3(c) and (d), the hump feature in Fig.~8(b), which lead to the
double peak $P(I)$ moves much faster to lower currents with
increasing microwave power. These distinct difference led us to
conclude that the double peak distributions in Fig. 3 were
originated from photon-assisted tunneling rather than the dynamic
bifurcation under strong ac driving described above.

However, as the intensity of ac field is increased the dominant
escape mechanism changes from PAT to Josephson bifurcation. Namely,
the onset of $A_{2}$ state results in double-peak distributions with
large inter-peak separation. In Fig.~9, the corresponding switching
current distributions $P(I)$ converted from the $\Gamma _{WKB}(I)$
presented in Fig.~8(b) are plotted. The existence of the double-peak
structure (curves labeled 2 and 3) can be clearly seen. Comparing
the results with the experimental data in Fig.~4 (namely the curves
with relative powers of 9, 9.8, 10, and 13 dB), we can see that the
shapes of $P(I)$ are quite similar and the applied power ratio
between curve 4 and curve 1 is $\sim 2.5$, also similar to the data
displayed in Fig.~4.

\subsection{Results at $\nu$ = 7.0 GHz and further discussions}

The results of 9.6 GHz microwave radiation suggest that the simple
classical model provides a reasonably good qualitative explanation
for the reappearance of the double-peak structure in Fig.~4 as ac
driving is intensified. Such explanation is expected to work also
for the data in Fig.~5 with 7-GHz microwave radiation.
Unfortunately, analysis based on Eq.~(8) does not work well for the
$\nu =7$ GHz data since the resonant peak appears below $1$ $\mu $A
(see Fig.~5). In this case, we can see from the dashed line in
Fig.~6 that $A_{2}$ becomes significantly greater than $\pi$/2,
which makes the expansion of the $\sin \phi $ and $\cos \phi $ terms
in obtaining Eq.~(8) inapplicable. \cite{rem3}

\begin{figure}[t]
\includegraphics[width=0.49\textwidth]{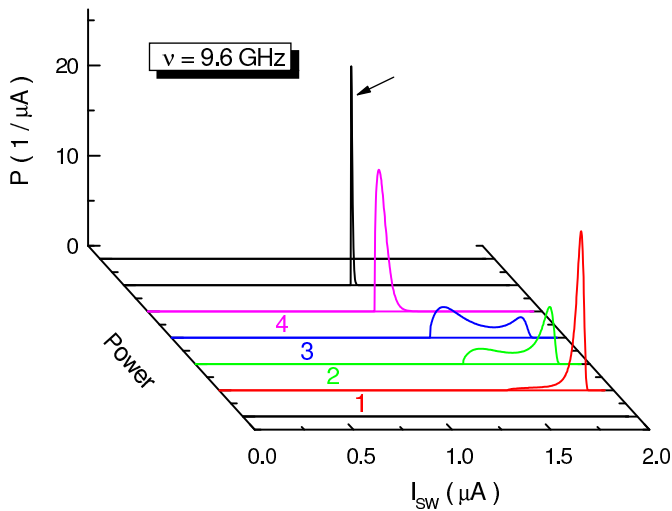}
\caption{(Color online) Switching current distributions $P(I)$
converted from $\Gamma_{WKB}$ in Fig.~8(b). Curve labels have
one-to-one correspondence with those in Fig.~8. The data give a
qualitative explanation for the development of the second resonant
peak in Fig.~4 (curves with relative powers of 9, 9.8, 10, and 13
dB). The curve pointed by an arrow is converted from that with arrow
in Fig.~8(b), and is displayed with its magnitude divided by 5.}
\end{figure}

Comparing the experimental results in Fig.~4 and Fig.~5, there are
two clear changes of the bifurcation-induced resonant peaks as the
microwave frequency $\nu$ decreases. One is the increase of the
inter-peak separation of the double-peak distribution and the other
is the rapid narrowing of the resonant peak width. The first change
can be easily understood from Fig.~6 that for fixed bias current and
power, $A_{2}$ is larger when $\nu$ is lower (compare the solid and
dashed lines at $I$ $=1.3$ $\mu $A). This directly leads to the
increase of the inter-peak separation for smaller $\nu $. It is
important to note that such trend is basically opposite to that
reported in the previous work where $W$ $\propto $ $i_{mw}^{2}$ is
smaller and bifurcation is not involved. \cite{jensen04,jensen06}

The rapid narrowing of the peak can also be explained by our model.
To demonstrate this, we plot in Fig.~8(b) the result with $\nu$ =
9.4 GHz (the top curve indicated by the arrow), a value smaller than
9.6 GHz used in the figure but not far away so that the
approximation used for Eq.~(8) is still valid. We have raised
$i^2_{mw}$ $\propto$ $W$ slightly so that the vertical part of the
curve sits at the same bias current as that of curve 4. We can see
that the calculated $\Gamma_{WKB}$ increases quickly, roughly by one
order of magnitude when $\nu$ is decreased only by 0.2 GHz. Such
increase would influence the peak width significantly since, unlike
the usual case such as those depicted in Figs.~3(c) and (d) where
escape rate changes continuously, the bifurcation related escape
rate versus dc bias current makes a sudden jump (dashed parts of the
curves in the figure) when $A_2$ appears. While the rate of escape
from the smaller $A_1$ amplitude oscillation is negligible, it
increases rapidly when the $A_2$ oscillation appears, which results
in switching to occur in a very small range of bias current and
therefore a much narrower resonant peak. Apparently, the lower the
$\nu$ is, the higher level the escape rate will suddenly jump to,
and the narrower the resonant peak will be. The curve pointed by the
arrow in Fig.~9 is the resulting distribution that appears
considerably sharper than curve 4 in the same figure.

It should be pointed out that from our simple model the ``primary"
and ``resonant" peaks in Fig.~9 have different meanings as
conventionally understood from the data such as those in Fig.~3.
According to the discussion on the results presented in Fig.~8, all
peaks in Fig.~9 are caused by tunneling from the $A_{2}$ state of
the resonantly driven phase particle. The double-peak structure is a
result of combined effect of non-trivial bias current dependence of
the potential well depth $U(i,\pi -\varphi _{0})-U(i,\varphi _{0})$
and the amplitude $A_{2}(i,W)$ at a given $W$. According to Eqs.~(1)
and (11), both quantities decrease as $i$ increases but at different
rates. This gives rise to the local minima and humps in Fig.~8, and
the ``resonant" peak in Fig.~9.

We note that quantitatively the experimental data and calculation
based on Josephson bifurcation do not agree well, thus further
improvements of the model are necessary. Among them, including
higher order terms in the expansion of $\sin \phi $ and $\cos \phi $
should be considered for in order to describe the data taken at $\nu
=7.0$ GHz. \cite{rem3} We emphasize that Josephson bifurcation is
used to explain the appearance of the second resonant peak in Fig.~4
for data taken at $9.8$ dB and $10$ dB. The double-peak structure at
much weaker ac driving, from about $0$ dB to $3$ dB, is due to the
quantum resonant phase escape process accompanied by barrier
suppression, \cite{fistul03,fistul00} which is not considered in the
present model.

\section{Summary}

We presented a systematic study of resonant phase escape in tilted
washboard potential of a Josephson junction (a dc SQUID with small
loop inductance) at $25$ mK in the presence of strong ac driving
(microwave radiation). The device was well-characterized using the
TA and MQT theories in the absence of the microwave field. For weak
ac driving $i_{mw}\ll 1$, we observed results that are interpreted
by quantum resonant escape of the phase involving single- and
two-photon absorptions with a suppressed potential barrier. At
larger $i_{mw}$, a second resonant peak, which is well separated
from the first one, would appear. The peak could locate at very low
bias current depending on the power and frequency of the microwave
applied. We proposed a model based on the classical equation of
motion to interpret the data. Our results indicated that at large
$i_{mw}$, the phase particle enters a bistable state due to
bifurcation of the nonlinear system and the oscillation state with
larger amplitude leads to the resonant peaks in the switching
current distributions locating far below the system's critical
current. These results are useful for the further studies of the
nonlinear response of Josephson junctions to strong ac driving and
for the detection of the qubit quantum states using a dc SQUID. In
addition, the results and analysis showed that for junctions with
strong ac driving, even at $T\ll T_{cr}$, the quantum-to-classical
crossover temperature, escape of the phase particle could be
dominated by classical processes. Our work therefore provides
further evidence of the dual character, classical and quantum
mechanical, of the Josephson junction system at $T\ll T_{cr}$ in an
ac driving field.

\section*{ACKNOWLEDGMENTS}

We thank V. Patel, W. Chen, and J. E. Lukens for providing us with
the Nb samples used in this work. The work at the Institute of
Physics was supported by the National Natural Science Foundation of
China (Grant Nos. 10534060 and 10874231) and the Ministry of Science
and Technology of China (Grant Nos. 2006CB601007, 2006CB921107, and
2009CB929102). S. Han was supported in part by NSF Grant No.
DMR-0325551.



\begin{thebibliography}{99}
\bibitem{rem} $C$ was mainly contributed by a designed parallel capacitance
in the device.
\bibitem{rem1} We point out that classical analysis should also work
though quantum-mechanical approach is used here. See
Refs.~\onlinecite{jensen04}, \onlinecite{jensen06}, and
\onlinecite{sun08} for the resonant peak development and
Refs.~\onlinecite{jensen04a} and \onlinecite{jensen06} for the
leftward shift of the primary peak.
\bibitem{rem2} Heating effect induced by microwave field was negligible
in our experiment. If we consider the ac current magnitude to be
less than $I_c$, the maximum heating power is about 0.6 nW. This
value is much smaller than that generated by junction's switching to
the normal state ($\sim$ 6 nW), which already shows negligible
influence since the experimental $T_{cr}$ agrees well with the
theory.
\bibitem{rem3} Such expansion is strict as $\phi$ $\rightarrow$ 0, and is
reasonable when $\phi$ $<$ $\pi$/2 $\sim$ 1.5. From the results in
Fig.~6, the treatment can be safely used for the 9.6 GHz data in
Fig.~4, but can be inaccurate for those in Fig.~5 at lower microwave
frequency.
\bibitem{rem4} In the corresponding quantum picture, a particle excited to
an energy level near the barrier top by multi-photon process has a
smaller probability to MQT out of the potential well than to decay
into lower energy levels in the well.
\end{thebibliography}
\end{document}